\documentclass[aps,prb,twocolumn,showpacs,10pt,floatfix,longbibliography]{revtex4-2}

\usepackage{graphicx}
\usepackage{subfigure}
\usepackage{amsmath}
\usepackage{amsfonts}

\usepackage{mathrsfs}
\usepackage{bbm}
\usepackage{subfigure}
\usepackage{xcolor}
\usepackage[colorlinks=true, urlcolor=blue, linkcolor=blue, citecolor=blue]{hyperref}

\usepackage[T1]{fontenc}

\usepackage[per-mode=symbol]{siunitx}
\usepackage{bm}
\usepackage[version=4]{mhchem} 

\usepackage{stackengine}[2013-09-11]
\newcommand\slashzero{\stackinset{c}{}{c}{}{/}{0}}

\usepackage{float}

\usepackage{dsfont}

\begin{document}

\title{Light-induced Andreev phase coherence and tunneling Hall effect in semi-Dirac systems}

\author{W.~Zeng}
\email{zeng@ujs.edu.cn}
\affiliation{Department of Physics, Jiangsu University, Zhenjiang 212013, China}

\begin{abstract}

We theoretically investigate the charge transport in a normal metal/normal metal/superconductor junction based on semi-Dirac materials. It is shown that off-resonant circularly polarized light applied to the central normal region induces an additional phase for the backreflected states. This light-induced phase depends on the electron's transverse momenta and becomes coherent via multiple reflections, leading to a transversely asymmetric Andreev reflection, which in turn produces a tunneling Hall effect. Both the longitudinal and transverse conductances are obtained within the nonequilibrium Green's function formalism. While the longitudinal conductance is insensitive to the light handedness and only acquires a finite phase shift with varying intensity, the transverse conductance reverses sign upon switching the handedness, indicating the reversal of the tunneling Hall current. Our results establish a phase-coherence mechanism for generating tunneling Hall currents in superconducting tunnel junctions, suggesting potential applications in superconducting electronics.

\end{abstract}
\maketitle

\section{Introduction}\label{intro}

In contrast to the isotropic linear dispersion near the Dirac cone in Dirac materials \cite{PhysRevLett.115.126803,PhysRevB.86.045443,RevModPhys.80.1337}, semi-Dirac materials (SDMs) host quasiparticles exhibiting linear dispersion along one momentum direction and quadratic dispersion along the orthogonal direction \cite{PhysRevX.14.041057,PhysRevLett.103.016402,PhysRevB.92.161115}. Such energy dispersion has been reported in a variety of systems and models, including $\ce{TiO2\slash VO2}$ superlattice \cite{PhysRevLett.102.166803,PhysRevB.81.035111}, strained or doped phosphorene \cite{PhysRevLett.112.176801,PhysRevLett.113.046804}, strained organic salt \cite{PhysRevB.74.033413,PhysRevB.84.075450}, and deformed graphene \cite{PhysRevB.80.153412}. In SDMs, the anisotropic band touching typically emerges at the critical point where two Dirac nodes merge, giving rise to many intriguing phenomena, such as the anisotropic transport of polaritons \cite{PhysRevLett.125.186601}, Landau-Zener oscillations \cite{PhysRevB.96.045424}, anisotropic tunneling \cite{Choi_2021,HUANG2023128671,10.1063/5.0147268}, orientation-dependent Andreev reflections \cite{Li_2022}, and the Goos-H\"anchen shift \cite{PhysRevB.109.035432}.

In conventional Dirac materials such as graphene and silicene, a Haldane-like mass gap can be produced by illuminating with circularly polarized light \cite{PhysRevB.84.235108,PhysRevLett.107.216601}, leading to numerous interesting phenomena in light-modulated superconducting transport, including valley polarization and $0$-$\pi$ transitions \cite{PhysRevB.89.064501,PhysRevB.110.155420,PhysRevB.105.094510,PhysRevB.105.064503}. However, in SDMs, such a gap cannot be opened by a circularly polarized light \cite{PhysRevB.91.205445,PhysRevB.109.235416,dyqm-xqbn,8c1n-8jh8,PhysRevB.99.075415}, implying that SDMs exhibit fundamentally different light-controlled transport behavior. Although the effects of light on the normal-state transport in SDMs have been extensively studied \cite{PhysRevB.98.235424,PhysRevB.99.075415,HUANG2023128671,PhysRevB.94.081103,PhysRevB.97.035422}, their superconducting transport properties remain largely unexplored. Moreover, it has been predicted that superconductivity can arise in two-dimensional SDMs with arbitrarily weak attraction in the presence of a random chemical potential \cite{Wang_2019}, while mean-field and renormalization group analyses suggest that $s$-wave pairing is energetically favored \cite{PhysRevB.96.220503,PhysRevB.100.155101}. This provides a foundation for studying the superconducting transport properties of semi-Dirac fermions.

As a fundamental phenomenon in superconducting transport, Andreev reflection \cite{PhysRevLett.74.1657,PhysRevLett.84.1804,FURUSAKI1991299} occurs at the normal metal/superconductor (NS) interface, converting incident electrons into holes while injecting Cooper pairs into the superconductor. This electron-hole conversion plays a key role in determining the conductance, phase coherence, and bound-state formation in NS and SNS (Josephson) junctions. Typically, Andreev reflection is symmetric in the transverse direction of the junction (perpendicular to the applied bias), which is responsible for the absence of the transverse tunneling current. However, it has been reported that Rashba and Dresselhaus spin-orbit coupling at the NS interface can break this symmetry \cite{PhysRevB.111.054512,PhysRevB.100.060507}, generating a finite transverse current and a tunneling Hall effect \cite{PhysRevLett.115.056602,PhysRevLett.117.166806,PhysRevLett.131.246301,PhysRevLett.134.176002}. In addition, phase-coherent Andreev reflections can take place in NNS or SNS double-junction tunnel structures, enabling the detection of specular Andreev reflection \cite{PhysRevB.103.144518}, the implementation of Andreev interferometers \cite{mcpw-z4yr}, and the realization of Josephson diode \cite{PhysRevB.109.184513}.

In this work, we propose a phase-coherent Andreev reflection induced by circularly polarized light in a semi-Dirac system and demonstrate that it can further generate a transverse tunneling current, thereby realizing the tunneling Hall effect. We consider the NNS double-junction structure based on SDMs. It is shown that the reflected state in the central normal region acquires an additional phase when the off-resonant circularly polarized light irradiation is applied to this region. The coherence of this light-induced phase gives rise to a transversely asymmetric Andreev reflection in the left normal lead, which is responsible for a finite tunneling transverse current. The longitudinal and transverse conductances are computed using the nonequilibrium Green's function (NEGF) formalism. While the longitudinal conductance is insensitive to the light handedness and only acquires a finite phase shift with varying intensity, the transverse conductance reverses sign upon switching the handedness, indicating the reversal of the tunneling Hall current. Our results establish a phase-coherence mechanism for generating tunneling Hall currents in superconducting tunnel junctions and suggest potential applications in superconducting electronics.

The remainder of this paper is organized as follows. The model Hamiltonian is explained in detail in Sec.\ \ref{mod}. Numerical calculations of the Andreev reflection coefficients and analysis of the phase-coherence mechanism are presented in Sec.\ \ref{aph}. Calculations of the transverse tunneling current and transverse conductance, together with an analysis of the tunneling Hall effect, are presented in Sec.\ \ref{sec:transverse_current_and_tunneling_hall_effect}. Finally, the conclusions are summarized in Sec.\ \ref{conc}.

\begin{figure}[tp]
\begin{center}
\includegraphics[clip = true, width =\columnwidth]{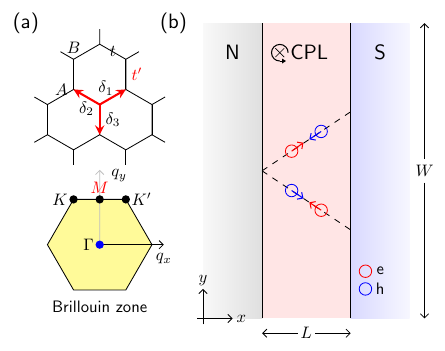}
\end{center}
\caption{ (a) Schematic illustrations of the honeycomb lattice and the first Brillouin zone. (b) Schematic of a light-irradiated tunnel junction. The central normal region ($0<x<L$) is irradiated by the off-resonant circularly polarized light (CPL). The superconducting region (S) occupies $x>L$, while the left normal lead (N) occupies $x<0$. The width of the junction is $W$. }
\label{fig:junction}
\end{figure}

\section{Model Hamiltonian}\label{mod}

The SDMs can be described by the tight-binding model with different values of nearest-neighbor hopping parameters \cite{PhysRevB.98.235424,PhysRevB.97.035422}
\begin{align}
    \hat{H}=-\mu\sum_{i,\sigma}(\hat{a}^\dagger_{i\sigma}\hat{a}_{i\sigma}+\hat{b}^\dagger_{i\sigma}\hat{b}_{i\sigma})+\sum_{\langle i,j\rangle,\sigma}(t_{ij}\hat{a}^\dagger_{i\sigma}\hat{b}_{j\sigma}+h.c.),
\end{align}
where $\mu$ is the chemical potential, $\hat{a}_{i\sigma}^\dagger$ ($\hat{b}_{i\sigma}$) denotes the creation (annihilation) operator for an electron on the $A$ ($B$) sublattice at site $i$ with spin $\sigma$. $\langle i,j\rangle$ runs over all nearest-neighbor hopping sites. $t_{ij}=t$ along the $\bm{\delta}_{1,2}$ directions, whereas $t_{ij}=t'$ along the $\bm{\delta}_3$ direction; see Fig.\ \ref{fig:junction}(a) (top panel). Here $\bm\delta_1=(\frac{\sqrt{3}a}{2},\frac{a}{2})$, $\bm\delta_2=(-\frac{\sqrt{3}a}{2},\frac{a}{2})$, and $\bm\delta_3=(0,-a)$. In momentum space, the Hamiltonian takes the form $\hat{H}=\sum_{\bm{q},\sigma}\hat{c}^\dagger_{\bm{q}\sigma}\hat{h}(\bm{q})\hat{c}_{\bm{q}\sigma}$, where $\hat{c}_{\bm{q}\sigma}=(\hat{a}_{\bm{q}\sigma},\hat{b}_{\bm{q}\sigma})^T$ and 
\begin{align}
    \hat{h}(\bm{q})=\begin{pmatrix}
        -\mu&f_{\bm{q}}\\
        f^*_{\bm{q}}&-\mu
    \end{pmatrix}
\end{align}
with $f_{\bm{q}}=te^{i\bm{q}\cdot\bm{\delta}_1}+te^{i\bm{q}\cdot\bm{\delta}_2}+t'e^{i\bm{q}\cdot\bm{\delta}_3}$. However, in the $\hat{c}_{\bm{q}\sigma}$ basis, the Hamiltonian is not of Bloch form since $f_{\bm{q}+\bm{G}} \neq f_{\bm{q}}$ \cite{BernevigHughes}, with $\bm{G}=\bm{b}_{1,2}$ denoting reciprocal lattice vectors. A gauge transformation on the $B$ sublattice, $b_{\bm{q}\sigma} \rightarrow e^{i\bm{q}\cdot\bm\delta_3} b_{\bm{q}\sigma}$, restores the Bloch form and yields $f_{\bm{q}} = t e^{i\bm{q}\cdot(\bm{\delta}_1-\bm{\delta}_3)} + t e^{i\bm{q}\cdot(\bm{\delta}_2-\bm{\delta}_3)} + t'$. The semi-Dirac phase appears when $t'=2t$, where the band dispersion is given by $\epsilon_{\bm{q}}=\pm|f_{\bm{q}}|-\mu$ with $\pm$ refers to the conduction band and valence band, respectively, and 
\begin{equation}
    |f_{\bm{q}}|
    =t\sqrt{2\cos(\sqrt{3}q_xa)+8\cos(\frac{\sqrt{3}q_xa}{2})\cos(\frac{3q_ya}{2})+6},
\end{equation}
The conduction and valence bands touch at $(q_x,q_y)=(0,\frac{2\pi}{3a})$, labeled as $M$ in the Brillouin zone [see Fig.\ \ref{fig:junction}(a) (bottom panel)]. Near $M$, the dispersion is quadratic along $x$ ($K\to M\to K'$) and linear along $y$ ($\Gamma \to M$). Focusing on the low-energy region around the Fermi surface, we expand about $M$ by setting $q_x=k_x$ and $q_y=2\pi/3a+k_y$, with $k_xa\ll1$ and $k_ya\ll1$, yielding
\begin{align}
    f_{M+\bm{k}}\simeq t(\frac{3}{4}k_x^2a^2-3ik_ya)= \alpha k_x^2-i\beta k_y,
\end{align}
where $\alpha=3a^2t/4$ and $\beta=3at$. Consequently, in the basis $\hat{c}_{\bm{k}\sigma}=(\hat{a}_{\bm{k}\sigma},e^{-i2\pi/3}\hat{b}_{\bm{k}\sigma})^T$, the low-energy Hamiltonian takes the form $\hat{H}=\sum_{\bm{k}\sigma}\hat{c}^\dagger_{\bm{k}\sigma}\hat{\mathcal{H}}(\bm{k})\hat{c}_{\bm{k}\sigma}$ with 
\begin{align}
    \hat{\mathcal{H}}(\bm{k})=\alpha k_x^2\hat{\tau}_x+\beta k_y\hat{\tau}_y-\mu,
\end{align}
where $\hat{\tau}_{x}$ and $\hat{\tau}_{y}$ are the Pauli matrices in pesudospin (sublattice) space.

The two-dimensional SDM-based tunnel junction is schematically shown in Fig.\ \ref{fig:junction}(b), with the longitudinal direction along the $x$-axis. The normal region occupies $x<L$, while the superconducting electrode covers $x>L$, where conventional $s$-wave pairing is assumed. The width of the junction is $W$. In the normal side, the off-resonant circularly polarized light is uniformly applied to a central region $0<x<L$, described by the time-dependent vector potential $\bm{A}(t)=A(\eta\sin\omega t,\cos\omega t)$, where $\eta=\pm1$ corresponds to right- and left-handed circularly polarized light, and $A$ and $\omega$ denote the light amplitude and frequency, respectively. The response of the electrons in the irradiated region can be obtained by the Peiere's substitution $\bm{k}\rightarrow \bm{k}+e\bm{A}(t)$, resulting in $\hat{\mathcal{H}}(\bm{k},t)=\hat{\mathcal{H}}(\bm{k})+\hat{\mathcal{H}}'(t)$, where
\begin{align}
    \hat{\mathcal{H}}'(t)&=\alpha[2\eta eAk_x\sin\omega t+(eA\sin\omega t)^2]\hat{\tau}_x\nonumber\\
    &\quad+\beta eA\cos(\omega t)\hat{\tau}_y.
\end{align}
By using Floquet theory, the effective Hamiltonian in the condition of high frequency ($\hbar \omega\gg evA$) is modified as \cite{PhysRevB.91.205445,PhysRevB.98.235424,PhysRevB.97.035422}
\begin{align}
    \hat{\mathcal{H}}_{\mathrm{eff}}(\bm{k})=\hat{\mathcal{H}}(\bm{k})+\frac{[\hat{\mathcal{H}}_{-1},\hat{\mathcal{H}}_{+1}]}{\hbar \omega}+\mathcal{O}(\frac{1}{\omega^2}),\label{eq:hf}
\end{align}
where $\hat{\mathcal{H}}_{\pm1}=\frac{1}{T}\int_{0}^T\mathcal{H}'(t)e^{\pm i\omega t}dt$ are the Fourier components of the time-dependent Hamiltonian with $T=2\pi/\omega$ being the time period of the light. After completing the calculation of the commutator in Eq.\ (\ref{eq:hf}), the effective Hamiltonian of the irradiated SDMs is obtained as 
\begin{align}
    \hat{\mathcal{H}}_{\mathrm{eff}}(\bm{k})=\hat{\mathcal{H}}(\bm{k})+\lambda k_x\hat{\tau}_z,
\end{align}
where $\lambda=2\eta \alpha\beta(eA)^2k_x/\hbar\omega$ is the illumination parameter. The chemical potential is taken as $\mu_S$, $\mu_C$, and $\mu_L$ in the superconducting region, central irradiated region, and left normal lead, respectively. In terms of the Nambu spinor $\hat{\psi}_{\sigma}=(\hat{c}_{\sigma},\hat{c}^\dagger_{\bar\sigma})^T$, the Bogoliubov-de Gennes (BdG) Hamiltonian can be written as $\hat{H}_{BdG}=\frac{1}{2}\sum_\sigma\int dx\hat{\psi}_{\sigma}^\dagger(x)\hat{\mathcal{H}}^\sigma_{BdG}(x)\hat{\psi}_{\sigma}(x)$, where
\begin{align}
    \hat{\mathcal{H}}^\sigma_{BdG}(x)=\begin{pmatrix}
        \hat{\mathcal{H}}-i\lambda(x)\hat{\tau}_z\partial_x&\hat{\Delta}(x)\\
        \hat{\Delta}^\dagger(x)&-\hat{\mathcal{H}}-i\lambda(x)\hat{\tau}_z\partial_x
    \end{pmatrix}.\label{eq:bdg}
\end{align}
Here $\lambda(x)=\lambda \Theta(x)\Theta(L-x)$ denotes the light-induced mass term in the central region, and $\Delta(x)=\Delta\Theta(x-L)$ is the $s$-wave pair potential, with $\Theta(x)$ being the Heaviside step function. Due to the spin degeneracy, it suffices to consider the spin-up sector of the BdG Hamiltonian in our model.

For $W\gg L$, we consider a periodic boundary along the $y$ direction, so that $k_y=2n\pi/W$ is a good quantum number with $n$ an integer, and the continuum Hamiltonian can be mapped onto a $k_y$-dependent one-dimensional chain with equal spacing $a$. For the normal side, the discrete Hamiltonian reads
\begin{align}
    \hat{H}_{N}&=\sum_{j,k_y}(\hat{c}^\dagger_{jk_y}\check{t}_{0}\hat{c}_{jk_y}+\hat{c}^\dagger_{jk_y}\check{t}_{x}\hat{c}_{j+1k_y}+h.c.).
\end{align}
The central irradiated region is assumed to extend over $1\leq j\leq N$ with a length $L=Na$, while the left (normal) and right (superconducting) leads cover $j\leq0$ and $j\geq N+1$, respectively. $\hat{t}_0$ is the $k_y$-contracted on-site energy matrix at the ($j$)th layer, with
\begin{align}
    \hat{t}_0=2t_\alpha\hat{\tau}_x+t_\beta\hat{\tau}_y\sin k_ya-\tilde{\mu},
\end{align}
where $t_\alpha=\alpha/a^2$, $t_\beta=\beta/a$. The chemical potential $\tilde{\mu}$ is given by $\tilde{\mu}=\mu_L$ for $j\leq 0$, $\tilde{\mu}=\mu_C$ for $1\leq j\leq N$, and $\tilde{\mu}=\mu_S$ for $j\geq N+1$. $\check{t}_x$ denotes the coupling between the ($j$)th and ($j+1$)th layers, with 
\begin{align}
    \check{t}_x=-t_\alpha\hat{\tau}_x-\frac{i\tilde{\lambda}}{2a}\hat{\tau}_z,
\end{align}
where $\tilde{\lambda}=0$ for $j\leq 0$ and $\tilde{\lambda}=\lambda$ for $1\leq j\leq N$. The junction Hamiltonian in the Nambu basis, with the right superconducting region included, is given
\begin{align}
    \hat{H}_{BdG}&=\sum_{j,k_y}(\hat{\psi}^\dagger_{jk_y}\check{H}_0\hat{\psi}_{jk_y}+\hat{\psi}^\dagger_{jk_y}\check{T}\hat{\psi}_{j+1k_y}+h.c.),
\end{align}
where the on-site energy matrix and the hopping matrix are given by
\begin{align}
    \check{H}_0=\begin{pmatrix}
        \hat{t}_0&\tilde\Delta\\
        \tilde\Delta&-\hat{t}_0
    \end{pmatrix},\quad \check{T}=\begin{pmatrix}
        \hat{t}_x&\slashzero\\
        \slashzero&-\hat{t}_x^\dagger
    \end{pmatrix},
\end{align}
respectively, with $\tilde\Delta=0$ for $j\leq N$ and $\tilde\Delta=\Delta$ for $j\geq N+1$.

\section{Phase coherent Andreev reflections}\label{aph}

\begin{figure}[tp]
\begin{center}
\includegraphics[clip = true, width =0.9\columnwidth]{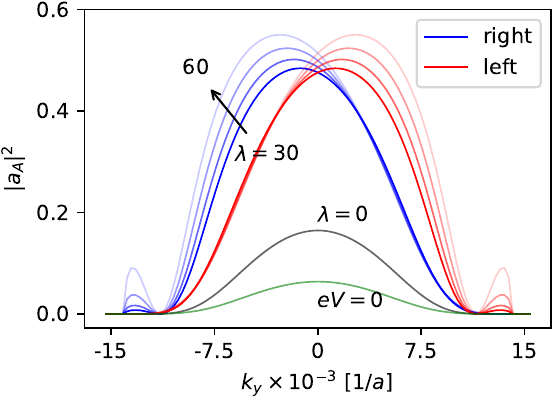}
\end{center}
\caption{$|a_A|^2$ vs $k_y$ for the central normal region irradiated by the right-handed circularly polarized light (blue), left-handed circularly polarized light (red), in the absence of the irradiated light (black), and at zero bias (green). The arrow marks the curves for the illumination parameter $\lambda$ varying from $30$ to $60$ in increments of $10$ (in units of $\Delta\cdot a$). The incident energy is set as $eV=0.8\Delta$ (black) and $eV=0.95\Delta$ (blue and red). The other parameters are $\mu_L=10\Delta$, $\mu_C=4\Delta$, $\mu_S=100\Delta$, $L=20a$, and $W=10000a$.}
\label{fig:ar}
\end{figure}

We choose the superconducting pair potential $\Delta=\SI{1}{\meV}$ as the energy unit and $a=\SI{1}{\angstrom}$ as the length unit throughout the numeric calculation. We consider the parameters $\alpha=\SI{0.75}{\electronvolt\angstrom^2}$ and $\beta=\SI{0.65}{\electronvolt\angstrom}$ corresponding to a typical SDM \cite{PhysRevB.94.081103}. In the experiment, the typical phonon energy is $\SI{0.25}{eV}$ with $eA=0.01\text{--}0.2\,\si{\per\angstrom}$, leading to the illumination parameter $|\lambda|\leq\SI{78}{\meV\angstrom}$, which therefore defines the parameter range of interest.

Within the framework of NEGF formalism, the net Andreev reflection coefficient can be obtained by \cite{RevModPhys.58.323,PhysRevB.59.3831}
\begin{align}
    |a_A|^2=\operatorname{tr}[\Gamma_{Le}\mathcal{G}^r_{eh}\Gamma_{Lh}\mathcal{G}^r_{he}].
\end{align}
Here the subscript $e$ ($h$) denotes the electron (hole) component in Nambu space. The retarded Green's function is given by the Dyson equation $\mathcal{G}^{r-1}=g^{r-1}-\Sigma^r$ with $g^r$ being the unperturbed Green's function of the central irradiated region and $\Sigma^r=\Sigma_{L}^r+\Sigma_{R}^r$ being the total retarded self-energy. $\Sigma_{L,R}^r$ due to coupling with the left and right leads can be calculated numerically by the recursive iteration method \cite{Sancho_1985,Sancho_1984}, and the level width matrices are given by the expression $\Gamma_{L,R}=-2\operatorname{Im}\Sigma^r_{L,R}$. The advanced Green's function is obtained by $\mathcal{G}^{a}=[\mathcal{G}^{r}]^\dagger$.

\begin{figure}[tp]
\begin{center}
\includegraphics[clip = true, width =\columnwidth]{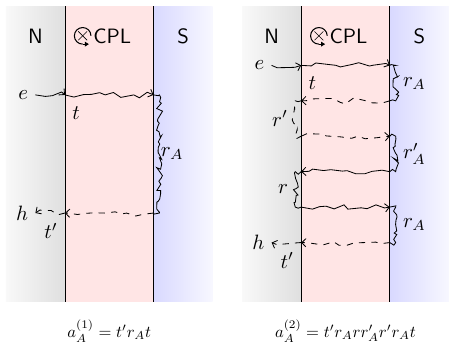}
\end{center}
\caption{Multiple-reflection processes in the central region of the junction: no normal reflections at the left boundary (left) versus two normal reflections at the left boundary (right). The solid and dashed lines denote electron and hole states, respectively. }
\label{fig:mutiar}
\end{figure}

The Andreev reflection coefficients are presented in Fig.\ \ref{fig:ar} at $eV=0.95\Delta$ with different illumination parameter $\lambda$. In the junction with a central region irradiated by right-handed circularly polarized light ($\lambda>0$), electrons with negative $k_y$ respond more strongly to the light, resulting in a considerable enhancement of the Andreev reflection as $\lambda$ increases (blue curves). By contrast, positive-$k_y$ electrons are largely unaffected, with the Andreev reflection coefficients changing only slightly as $\lambda$ increases. Consequently, the Andreev reflections exhibits an asymmetric behavior with respect to $k_y$. This asymmetry can also be induced by left-handed circularly polarized light ($\lambda< 0$), resulting in the $k_y$-resolved $|a_A|^2$ pattern skewed to the opposite direction (red curves), and it disappears when the light field is absent (black curve). The Andreev reflection coefficient remains unchanged under the simultaneous reversal of $k_y$ and $\lambda$, \textit{i.e.}, $|a_A(k_y,\lambda)|^2=|a_A(-k_y,-\lambda)|^2$.

The physical origin of the numerically obtained asymmetric Andreev reflection is attributed to the light-induced phase coherence in the central spacer. The net Andreev reflection amplitude in the left normal lead can be constructed as a geometric series of multiple reflections \cite{kulik2012quantum,datta1997electronic}
\begin{align}
    a_A&=t'r_At+t'r_Arr'_Ar'r_At+t'r_A(rr'_Ar'r_A)^2t+\cdots\nonumber\\
    &=t'r_A[1+rr'r_Ar_A'+(rr'r_Ar_A')^2+(rr'r_Ar_A')^3+\cdots]t\nonumber\\
    &=\frac{tt'r_A}{1-rr'r_Ar_A'},\label{eq:aA}
\end{align}
where $r$ ($r'$) and $t$ ($t'$) are the normal reflection and transmission amplitudes for the electron (hole) state at the left boundary of the central irradiated region, respectively, while $r_A$ ($r'_A$) is the Andreev reflection amplitude for $e\rightarrow h$ ($h\rightarrow e$) at the right boundary. The successive terms in this geometric series have a simple physical interpretation. The first term is the Andreev process without any normal reflections at the left boundary of the central region [Fig.\ \ref{fig:mutiar} (left)], the second term for Andreev process with two normal reflections at the left boundary [Fig.\ \ref{fig:mutiar} (right)], the third term for Andreev process with four normal reflections and so on. The net Andreev reflection coefficient can be obtained by squaring $a_A$ in Eq.\ (\ref{eq:aA})
\begin{align}
    |a_A|^2=\frac{|tt'r_A|^2}{1+|rr'r_Ar_A'|^2-2|rr'r_Ar_A'|\cos\theta},\label{eq:att}
\end{align}
where $\theta=\arg(rr'r_Ar_A')$ is the total phase of the reflection amplitudes.

In order to make the physical picture clear, we restrict our analytical analysis to the limit of small $\lambda$ and $k_y$ in a nearly zero-bias junction, \textit{i.e.}, $\lambda\ll\mu$, $k_y\rightarrow0$ and $E=eV\ll\Delta$, which is sufficient to capture certain qualitative properties of the total reflection phase $\theta$ and the Andreev reflection coefficient $|a_A|^2$. We note that no such restriction is applied in our numerical calculations. 

In this regime, the total reflection phase is finally obtained as (see Appendix A for details)
\begin{align}
    \theta\simeq-\pi+\sqrt{\mu}EL+\frac{\lambda k_y}{2\mu^{3/2}}.\label{eq:phase}
\end{align}
Apart from the conventional kinetic phase $\sqrt{\mu}EL$ arising from the motions of electrons and holes in the central spacer, an additional light-induced phase emerges [the third term in Eq.\ (\ref{eq:phase})]. Most interestingly, this additional phase is proportional to $k_y$, resulting in a $k_y$-dependent $\theta$ that exhibits an asymmetric behavior
\begin{align}
    \theta(k_y,\lambda)\neq\theta(-k_y,\lambda).\label{eq:ta1}
\end{align}
However, $\theta$ remains invariant under the simultaneous reversal of $k_y$ and $\lambda$
\begin{align}
    \theta(k_y,\lambda)=\theta(-k_y,-\lambda)\label{eq:ta2}
\end{align}
Although Eq.\ (\ref{eq:phase}) is derived in the limit of small $\lambda$, the parity properties of $\theta$ with respect to $k_y$ in Eqs.\ (\ref{eq:ta1}) and (\ref{eq:ta2}) remain valid even beyond this regime. Moreover, Eq.\ (\ref{eq:att}) shows that the Andreev reflection coefficient is an even function of $\theta$, as it depends on $\cos\theta$. Consequently, with the help of Eqs.\ (\ref{eq:ta1}-\ref{eq:ta2}), $|a_A|^2$ should be asymmetric with respect to $k_y$ with the same $\lambda$ but remains unchanged under the simultaneous reversal of $k_y$ and $\lambda$
\begin{gather}
    |a_A(k_y,\lambda)|^2\neq |a_A(-k_y,\lambda)|^2,\\
    |a_A(k_y,\lambda)|^2= |a_A(-k_y,-\lambda)|^2,\label{eq:sym}
\end{gather}
which account for the results obtained in our numerical calculations.

In addition, for a zero-bias junction, the skew Andreev reflection is always absent regardless of whether the central region is irradiated by light. This arises from the absence of the kinetic phase accumulated in this region, which follows straightforwardly from the analytical expression in Eq.\ (\ref{eq:phase}) in the small $\lambda$ regime: For $E=0$, the total phase $\theta$ is proportional to $-\pi+\lambda k_y$, the Andreev reflection coefficient (proportional to $\cos\theta$) is always even with respect to $k_y$. In fact, for the more general case beyond the small-$\lambda$ regime, the kinetic phase is proportional to $(k_e-k_h)L$ and thus vanishes at zero energy due to $k_e=k_h$ as a consequence of particle-hole symmetry, where $k_e$ and $k_h$ are the longitudinal wave vectors for the electron and hole states in the central region, respectively. Consequently, the kinetic phase is also absent even beyond the small-$\lambda$ regime, leading to the absence of the asymmetric Andreev reflection, consistent with the numerical results shown by the green curve in Fig.\ \ref{fig:ar} for $eV=0$ and $\lambda=30\Delta a$.

\begin{figure}[tp]
\begin{center}
\includegraphics[clip = true, width =0.9\columnwidth]{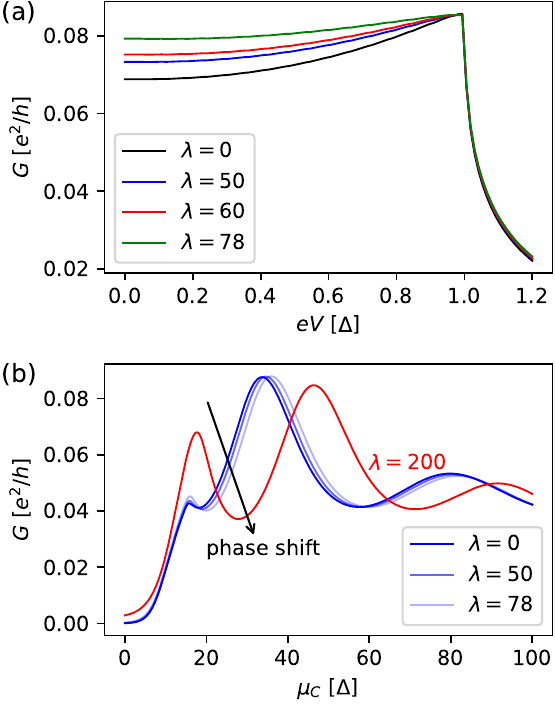}
\end{center}
\caption{$G$ as a function $eV$ (a) and as a function of $\mu_C$ (b). Parameters are $\mu_C=40\Delta$ in (a) and $eV=0.5\Delta$ in (b). Other parameters are the same as in Fig.\ \ref{fig:ar}.}
\label{fig:gl}
\end{figure}

The longitudinal conductance at zero temperature can be obtained by the Blonder-Tinkham-Klapwijk formula \cite{PhysRevB.25.4515}
\begin{align}
    G=\frac{e^2}{Wh}\sum_{k_y}(|t_Q|^2+2|a_A|^2),
\end{align}
where $|t_Q|^2=\operatorname{tr}[\Gamma_{Le}(\mathcal{G}^r\Gamma_R\mathcal{G}^a)_{ee}]$ is quasiparticle transmission coefficient. For the subgap regime ($eV\leq\Delta$), $G$ is determined by the Andreev reflection process. Because the calculation of $G$ involves summing over $k_y$ within a symmetric interval, it cannot distinguish between the left- and right-handed circularly polarized light under the symmetry of Eq.\ (\ref{eq:sym}). The longitudinal conductance spectrum of the junction irradiated by the right-handed circularly polarized light is presented in Fig.\ \ref{fig:gl}(a). It is shown that the subgap conductance increases with the increasing of $eV$, reaching a peak at $eV=\Delta$, and can be enhanced by increasing $\lambda$ in the experimentally achievable regime. In addition, $G$ exhibits a Fabry-P\'{e}rot oscallisiton behavior with the variation of the chemical potential of the central region ($\mu_C$). In the experimentally achievable $\lambda$ regime, a small phase shift of $G$ can be observed with increasing of $\lambda$, as the blue curves shown in Fig.\ \ref{fig:gl}(b), where $\lambda$ is set as $\lambda=0$, $50\Delta a$, and $78\Delta a$. This phase shift is attributed to the light-induced additional phase of $a_A$. We note that the light-induced phase increases with the transverse momentum $k_y$ and vanishes at $k_y = 0$. Since the longitudinal conductance is obtained by summing over all transverse modes and is dominated by contributions near $k_y = 0$, the overall effect of this additional phase on the longitudinal conductance is limited, resulting in only a small phase shift within the experimentally accessible $\lambda$ regime. For comparison, we also present in Fig.\ \ref{fig:gl}(b) the calculated curve for a larger $\lambda$, where a pronounced phase shift becomes evident.

\section{Transverse current and tunneling Hall effect}\label{sec:transverse_current_and_tunneling_hall_effect}
Most notably, the predicted phase-coherent skew Andreev reflection may induce a charge imbalance transverse to the applied bias, leading to a tunneling Hall effect \cite{PhysRevLett.115.056602,PhysRevLett.117.166806,PhysRevLett.131.246301,PhysRevLett.134.176002}. The transverse current carried by the $k_y$ mode can be evaluated via
\begin{align}
    I_y(k_y)&=\frac{-ie}{Wh}\int_{\mathbb{R}} d\omega\,v_y\operatorname{tr}[\hat{\tau}_z\hat{\sigma}_y\mathcal{G}^{<}_{ii}(\omega,k_y)],\label{eq:tj}
\end{align}
where $k_y=2n\pi/W$ with $n$ being an integer and $W$ being the width of the junction, $v_y(\varepsilon)=\partial \varepsilon_{k_y}/\partial k_y$ is the group velocity along the transverse direction. $\mathcal{G}_{ij}^<$ is the lesser Green's functions connecting the the ($i$)th and ($j$)th layers in nonequilibrium Green's function formalism, which can be obtained by the Keldysh equation in the energy domains
\begin{align}
    \mathcal{G}^<(\omega)=\mathcal{G}^r(\omega)\Sigma^<(\omega)\mathcal{G}^a(\omega).\label{eq:kel}
\end{align}
The lesser self-energy is given by $\Sigma^<=i\check{F}_L\Gamma_L+if_0\Gamma_R$, where $f_0=f(\omega)$ is the Fermi-Dirac distribution function and $\check{F}_L=\operatorname{diag}\{f_L,\tilde{f}_L\}$. The shorthand notations are defined as $f_L=f(\omega-eV)$ and $\tilde{f}_L=f(\omega+eV)$ with $V$ being the bias voltage. With the help of Eq.\ (\ref{eq:kel}), $I_y(k_y)$ can be reduced to $I_y(k_y)=I_1(k_y)+I_2(k_y)$, where 
\begin{align}
    I_1(k_y)&=\frac{e}{Wh}\int_{\mathbb{R}} d\omega\, v_{y}\operatorname{tr}[\hat{\sigma}_y\mathcal{G}^r_{eh}\Gamma_{Lh}\mathcal{G}^a_{he}(\tilde{f}_L-f_L)\nonumber\\
    &\quad+\hat{\sigma}_y(\mathcal{G}^r\Gamma_R\mathcal{G}^a)_{ee}(f_0-f_L)],\label{eq:I1}\\
    I_2(k_y)&=-\frac{e}{Wh}\int_{\mathbb{R}} d\omega\, v_{y}\operatorname{tr}[\hat{\sigma}_y\mathcal{G}^r_{he}\Gamma_{Le}\mathcal{G}^a_{eh}(f_L-\tilde{f}_L)\nonumber\\
    &\quad+\hat{\sigma}_y(\mathcal{G}^r\Gamma_R\mathcal{G}^a)_{hh}(f_0-\tilde{f}_L)].\label{eq:I2}
\end{align}
Here the subscript $e$ ($h$) denotes the electron (hole) component in Nambu electron-hole space. In Eqs.\ (\ref{eq:I1}-\ref{eq:I2}), the first terms denote the transverse transport induced by the Andreev reflections, whereas the second terms arise from quasiparticle tunneling and contribute only when the excitation energy exceeds the superconducting gap. The total transverse current is given by $I_y=\sum_{k_y}I_y(k_y)$, leading to and the transverse conductance $G_T=dI_y/dV$. Similar to the transverse current, $G_T$ can be decomposed into contributions from Andreev reflection and quasiparticle tunneling processes. At zero temperature, the transverse conductance arises from the Andreev process is obtained as
\begin{align}
    G_{T,A}(eV)&=-\frac{e^2t_\beta}{Wh}\sum_{k_y}\cos(k_ya)(A_e|_{eV}-A_e|_{-eV}\nonumber\\
    &\quad+A_h|_{eV}-A_h|_{-eV}),
\end{align}
with the coefficients $A_e=\operatorname{tr}(\hat{\sigma}_y\mathcal{G}^r_{eh}\Gamma_{Lhh}\mathcal{G}^a_{he})$ and $A_h=\operatorname{tr}(\hat{\sigma}_y\mathcal{G}^r_{he}\Gamma_{Lee}\mathcal{G}^a_{eh})$. The transverse conductance arising from quasiparticle tunneling above the superconducting gap is obtained as
\begin{align}
    G_{T,Q}(eV)=-\frac{e^2t_\beta}{Wh}\sum_{k_y}\cos(k_ya)(B_e|_{eV}-B_h|_{-eV}),
\end{align}
where the coefficients are $B_{e/h}=\operatorname{tr}[\hat{\sigma}_y(\mathcal{G}^r\Gamma_R\mathcal{G}^a)_{ee/hh}]$. The total transverse conductance is finally obtained by combining the Andreev and quasiparticle contributions
\begin{align}
    G_T=G_{T,A}+G_{T,Q}.
\end{align}

\begin{figure}[tp]
\begin{center}
\includegraphics[clip = true, width =0.9\columnwidth]{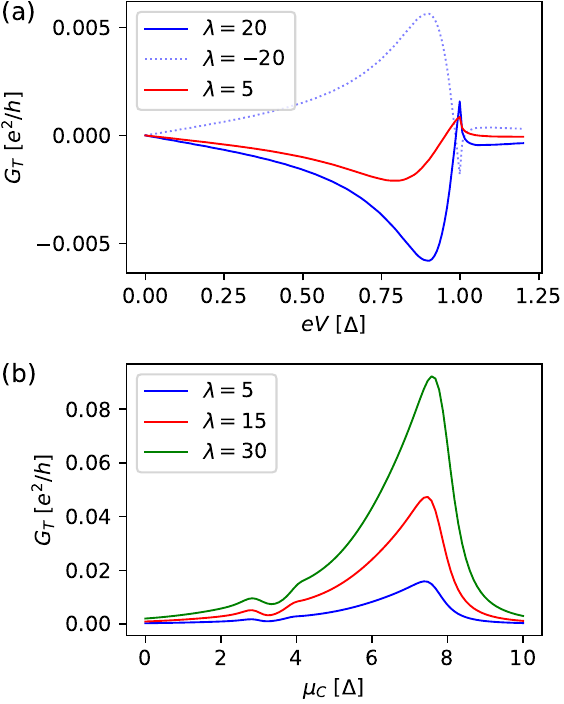}
\end{center}
\caption{$G_T$ as a function $eV$ (a) and as a function of $\mu_C$ (b). Parameters are $\mu_C=45\Delta$ in (a) and $eV=0.8\Delta$ in (b). Other parameters are the same as in Fig.\ \ref{fig:ar}.}
\label{fig:gt}
\end{figure}

The transverse conductance $G_T$ versus the incident energy $eV$ under right-handed circularly polarized light ($\lambda=20\Delta a$) is shown by the blue solid curve in Fig.\ \ref{fig:gt}(a). It is shown that $G_T$ is zero at $eV=0$ due to the particle-hole symmetry, where the Andreev reflection is symmetric with respect to the transverse momentum, indicating the absence of the transverse current. In the subgap regime, $G_T$ is governed by the Andreev process. As the incident energy increases, $G_T$ rises steadily, reaching a maximum at $eV\simeq0.85\Delta$, and then gradually decreases. A pronounced peak appears at $eV = \Delta$. A similar behavior persists under changes in the illumination parameters, as shown by the red curve in Fig.\ \ref{fig:gt}(a) for $\lambda=5\Delta a$. In contrast to the longitudinal conductance $G$, $G_T$ shows a fully opposite response under illumination with opposite circular polarizations. The blue dotted curve in Fig.\ \ref{fig:gt}(a) presents the transverse conductance versus $eV$ under left-handed circularly polarized light with the identical illumination parameter ($\lambda=-20\Delta a$). It is seen that, for a given $eV$, opposite circular polarization gives rise to transverse conductances with equal magnitude but opposite sign, implying transverse currents of the same strength flowing in opposite directions. $G_T$ versus the chemical potential of the central region $\mu_C$ at $\lambda=5\Delta a$ is shown by the blue curve in Fig.\ \ref{fig:gt}(b). Increasing the chemical potential of the central region produces two competing effects: it enhances $G_T$ by increasing the local density of states, but simultaneously suppresses $G_T$ due to the increased Fermi wavevector mismatch with the left normal lead. As a result, $G_T$ rises with $\mu_C$ and then decreases, reaching a maximum at $\mu_C\simeq7.5\Delta$. Increasing the light intensity produces a behavior similar to that described above, while the concomitant enhancement of the Andreev reflection asymmetry leads to a significant increase in $G_T$, as shown by the red and blue curves in Fig.\ \ref{fig:gt}(b) corresponding to $\lambda = 15\Delta a$ and $30\Delta a$, respectively.

\begin{figure}[tp]
\begin{center}
\includegraphics[clip = true, width =0.9\columnwidth]{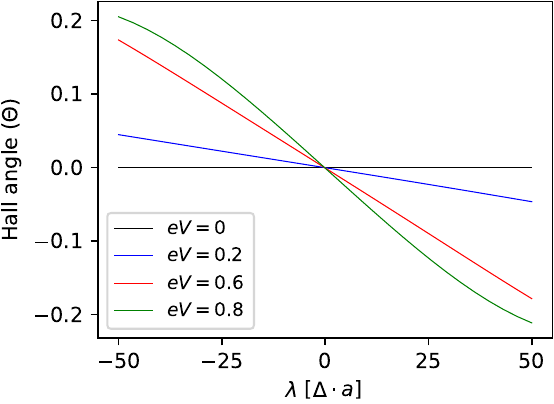}
\end{center}
\caption{$\Theta$ as a function of $\lambda$ with different $eV$ at $\mu_L=15\Delta$ and $\mu_C=40\Delta$. Other parameters are the same as in Fig.\ \ref{fig:ar}. }
\label{fig:hall}
\end{figure}

The efficiency of the tunneling Hall effect can be characterized by the Hall angle $\Theta=G_T/G$. The Hall angle as a function of the illumination parameter is presented in Fig.\ \ref{fig:hall} at different incident energy. It is shown that the Hall angle exhibits an odd dependence on $\lambda$, with an approximately linear scaling as $\lambda$ increases. At zero bias, the predicted tunneling Hall effect is absent regardless of whether illumination is applied. Under a finite bias, however, the Hall angle generally increases with increasing bias voltage.

We note that the tunneling Hall effect proposed in our model arises from a mechanism fundamentally different from those reported previously. In earlier works, transverse currents are generally induced by external fields or interface engineering that create anisotropic Fermi surfaces across the junction. The resulting transverse current arises from the skew tunneling caused by the Fermi surface mismatch. Specifically, this process generally involves additional intrinsic degrees of freedom, often accompanied by spin \cite{PhysRevLett.115.056602,PhysRevLett.117.166806,PhysRevB.111.054512,PhysRevB.100.060507}, valley \cite{PhysRevLett.131.246301}, or chirality \cite{PhysRevB.110.024511} Hall currents. By contrast, in our model the asymmetric tunneling originates solely from the light-induced phase coherence of multiple reflections, without requiring any additional degrees of freedom. As a result, a pure transverse charge current is generated.

\section{conclusions}\label{conc}
To conclude, we theoretically study the transport properties of a normal metal/normal metal/superconductor junction based on SDMs. We demonstrate that the off-resonant circularly polarized light applied to the central normal region of the junction induces an additional reflection phase for the carriers in this region, which depends on their transverse momenta. The coherence of this additional light-induced phase gives rise to a transversely asymmetric Andreev reflection, which is responsible for finite transverse current, leading to the tunneling Hall effect. The longitudinal and transverse conductances are further calculated using the NEGF formalism. It is shown that the longitudinal conductance is insensitive to the handedness of the circularly polarized light, but can exhibit a finite phase shift upon tuning the light intensity. In contrast, reversing the light handedness results in a sign change of the transverse conductance, indicating a corresponding reversal of the tunneling Hall current. Our results establish a phase-coherence mechanism for generating tunneling Hall currents in superconducting tunnel junctions, suggesting potential applications in superconducting electronics.

\section*{Acknowledgements}\label{ack}
This work was supported by the National Natural Science Foundation of China (Grant No.\ 12504052), the Natural Science Foundation of Jiangsu Province (Grant No.\ BK20250838), and the Natural Science Foundation of the Jiangsu Higher Education Institutions of China (Grant No.\ 25KJB140004).

\appendix

\section{Derivation of Eq.\ (\ref{eq:phase})}

We assume $\mu_L=\mu_C=\mu$ and adopt the heavy-doping limit in the superconducting region ($\mu_S\gg\mu$). Furthermore, we focus on small $\lambda$ and on the scattering states near the Fermi surface with small $k_y$, \textit{i.e.}, $\lambda\ll\mu$, $E\ll\Delta$, and $k_y\to0$.

We note that the above assumptions are introduced solely for the purpose of qualitative analysis, as they yield compact and analytically instructive description of the Andreev reflection phase, which can help us understand the main physics in our model. No such restrictions are imposed in the numerical calculations presented in the main text.

By introducing the characteristic energy $\epsilon_0=\beta^2/\alpha$ and length $\ell_0=\alpha/\beta$ to rescale the quantities $\bm{k}\ell_0\rightarrow \bm{k}$, $\mu/\epsilon_0\rightarrow\mu$, $\lambda/\epsilon_0\ell_0\rightarrow\lambda$, and $\Delta/\epsilon_0\rightarrow\Delta$, the dimensionless Hamiltonian in the normal region is given by
\begin{align}
    \mathcal{H}_{e/h}=\pm(k_x^2\sigma_x+k_y\sigma_y-\mu)+\lambda(x)k_x\sigma_z,\label{eq:heh}
\end{align}
where $\pm$ for the electron states ($\mathcal{H}_e$) and hole states ($\mathcal{H}_h$), respectively. For the given incident energy $E$ and the  conserved transverse wave number $k_y$, the propagating modes are given by
\begin{align}
    \chi_{e,\pm}(x)&=\begin{pmatrix}
        \varsigma_e^{(\pm)} e^{-i\phi_{e,-}/2}\\
        e^{i\phi_{e,-}/2}\\
        0\\
        0
    \end{pmatrix}e^{\pm ik_{e,-}x},\label{eq:eg}
    \\
    \chi_{h,\pm}(x)&=\begin{pmatrix}
    0\\0\\
        \varsigma_h^{(\pm)} e^{-i\phi_{h,-}/2}\\
        e^{i\phi_{h,-}/2}
    \end{pmatrix}e^{\pm ik_{h,-}x}
,\label{eq:pm}
\end{align}
where $\chi_{e/h,+(-)}$ indicates the propagating electron/hole state in the positive (negative) directions along the $x$-axis. The longitudinal wave number is given by
\begin{align}
	k_{e/h,-}=\sqrt{(\rho_{e/h}-\lambda^2)/2},
\end{align}
where $\rho_{e/h}=\sqrt{\lambda^4+4[(\mu\pm E)^2-k_y^2]}$. The parameters $\varsigma_{e/h}^\pm$ are given by $\varsigma_{e/h}^-=1/\varsigma_{e/h}^+$ with
\begin{align}
	\varsigma_{e/h}^+=\sqrt{\frac{\mu+\varrho_{e/h} E+\lambda k_{e/h,-}}{\mu+\varrho_{e/h} E-\lambda k_{e/h,-}}},
\end{align}
where $\varrho_{e}=1$ and $\varrho_{h}=-1$. The evanescent modes are given by
\begin{align}
    \varphi_{e,\pm}(x)&=\begin{pmatrix}
        -e^{i(\phi_{e,+}\pm\phi_{e,F})/2}\\
        e^{-i(\phi_{e,+}\pm\phi_{e,F})/2}\\
        0\\0
    \end{pmatrix}e^{\mp k_{e,+}x},\\
    \varphi_{h,\pm}(x)&=\begin{pmatrix}
        0\\0\\
        -e^{i(\phi_{h,+}\mp\phi_{h,F})/2}\\
        e^{-i(\phi_{e,+}\mp\phi_{h,F})/2}
    \end{pmatrix}e^{\mp k_{h,+}x},\label{eq:eg2}
\end{align}
where $\varphi_{e/h,\pm}$ indicate the states decaying as $x\rightarrow\pm\infty$, $k_{e/h,+}=\sqrt{(\rho_{e/h}+\lambda^2)/2}$ are real and positive. The angle parameters $\phi_{e/h,\pm}$ and $\phi_{e/h,F}$ in Eqs.\ (\ref{eq:eg}-\ref{eq:eg2}) are given by
\begin{gather}
    \sin(\phi_{e/h,\pm})=\frac{k_y}{\sqrt{(\mu\pm E)^2\pm\lambda^2k_{e/h,\pm}^2 }},\\
     \sin(\phi_{e/h,F})=\frac{\lambda k_{e/h,+}}{\sqrt{(\mu\pm E)^2+\lambda^2k_{e/h,+}^2 }}.
\end{gather}
It should be noted that the angle parameters $\phi_{e/h,\pm}$ and $\phi_{e/h,F}$ exhibit different parity behaviors with respect to $k_y$. $\phi_{e/h,\pm}$ is odd under $k_y\rightarrow-k_y$ while $\phi_{e/h,F}$ is even.

In the superconducting region, the four propagating states along the positive direction (or decaying as $x\to\infty$) are given by 
\begin{align}
    \chi_{s,\pm}=\begin{pmatrix}
        e^{\pm i\gamma/2}\\e^{\pm i\gamma/2}\\e^{\mp i\gamma/2}\\e^{\mp i\gamma/2}
    \end{pmatrix}e^{\pm iq_1x},\quad
\varphi_{s,\pm}=\begin{pmatrix}
        e^{\mp i\gamma/2}\\-e^{\pm i\gamma/2}\\e^{\pm i\gamma/2}\\-e^{\pm i\gamma/2}
    \end{pmatrix}e^{\pm iq_2x},
\end{align}
where $\gamma=\arccos(E/\Delta)$, $q_{1}=\sqrt{\mu_S\pm i\Omega}$, and $q_2=\sqrt{-\mu_S\pm i\Omega}$ with $\Omega=\sqrt{\Delta^2-E^2}$.

To determine the reflection phase of $r$, we consider the following scattering wave function
\begin{align}
    \Psi(x)=\left\{
        \begin{array}{ll}
         \chi_{e,-}+r\chi_{e,+}+r_d\varphi_{e,+}, & 0<x<L, \\
          t\chi^0_{e,-}+t_d\varphi^0_{e-}, & x\leq 0,
        \end{array}
    \right.\label{eq:gf0}
\end{align}
where $r$ and $t$ are the reflection and transmission amplitudes, respectively, while $r_d$ and $t_d$ are the corresponding amplitudes for the evanescent modes. $\chi_e^0$ and $\varphi_e^0$ are the scattering modes of the left normal lead in the absence of the light irradiation, which can be obtained by setting $\lambda=0$ in $\chi_e$ and $\varphi_e$, respectively. 

The continuity of $\Psi(x)$ at the boundary yields
\begin{align}
	\Psi(0^-)=\Psi(0^+).\label{eq:bc1}
\end{align}
Another boundary condition can be derived by antisymmetrizing the light-induced mass term in Eq.\ (\ref{eq:heh})
\begin{align}
    \lambda \hat{\tau}_z k_x\Theta(x)\rightarrow\frac{\lambda\hat{\tau}_z}{2}\{k_x,\Theta(x)\},
\end{align}
which ensures the hermiticity of the Hamilton operator. Integrating the eigenvalue function of $\Psi$ over $[-\varepsilon,\varepsilon]$ with $\varepsilon\rightarrow0^+$ yields
\begin{align}
    \partial_x\Psi(0^+)-\partial_x\Psi(0^-)=-\frac{\lambda\hat{\nu}_z\hat{\tau}_y}{2}\Psi(0),\label{eq:bc2}
\end{align}
where $\hat{\nu}_z$ is the Pauli matrix in Nambu space. The scattering amplitudes can be obtained by matching $\Psi(x)$ at $x=0$ according to the boundary conditions given in Eqs.\ (\ref{eq:bc1}) and (\ref{eq:bc2}). The reflection amplitude of $r$ is given by $\theta_{r}=\arg(r)$. By substituting $\chi_e \to \chi_h$ and $\phi_e \to \phi_h$ into Eq.\ (\ref{eq:gf0}) and repeating the same procedure, the reflection phase of $r'$ is obtained as $\theta_{r'} = \arg(r')$. After some algebra, one finds 
\begin{gather}
    \tan\theta_r=1-\frac{\sin^2(\frac{\phi_F}{2})+2\cos^2(\frac{\phi_F}{2}+\phi_e)}{4\cos(\frac{\phi_F}{2}+\phi_e)\sin(\frac{\phi_F}{2})}\frac{\lambda}{k_e},\\
    \tan\theta_{r'}=-1+\frac{\sin^2(\frac{\phi_F}{2})+2\cos^2(\frac{\phi_F}{2}-\phi_h)}{4\cos(\frac{\phi_F}{2}-\phi_h)\sin(\frac{\phi_F}{2})}\frac{\lambda}{k_h},
\end{gather}
where $\phi_F=\lambda/\sqrt{\mu}$, $q_{e/h}=\sqrt{\mu\pm E}$, and $\phi_{e/h}=k_y/(\mu\pm E)$. Consequently, the total reflection phase of $r$ and $r'$ is given by 
\begin{align}
     \theta_r+\theta_{r'}\simeq\frac{\lambda k_y}{2\mu^{3/2}}.
 \end{align} 

The reflection phase of $r_A$ can be determined by considering the following scatting wave function
\begin{widetext} 
\begin{align}
    \Psi(x)=\left\{
        \begin{array}{ll}
         \chi_{e,+}+r_A\chi_{h,-}+r_{A,d}\varphi_{h,-}+r_{N}\chi_{e,-}+r_{N,d}\varphi_{e,-}, & 0<x<L, \\
          c_1\chi_{s,+}+c_2\chi_{s,-}+c_3\varphi_{s,+}+c_4\varphi_{s,-}, & x\geq L,
        \end{array}
    \right.\label{eq:gf1}
\end{align}
where $r_A$ and $r_{A,d}$ are the Andreev reflection amplitudes for the propagating and evanescent modes, respectively, while $r_{N}$ and $r_{N,d}$ are the normal reflection amplitudes for the propagating and evanescent modes, respectively. $c_{1\!-\!4}$ are the transmission coefficients in the superconducting region. Matching the wave function at $x=L$ gives rise to $r_A$, and then the reflection phase of $r_A$ is obtained by $\theta_{r_A}=\arg(r_A)$.
\end{widetext}
Similarly, $r'_A$ can be obtained by Substituting $\chi_e\to\chi_h$ and $\varphi_e\to\varphi_h$ in Eq.\ (\ref{eq:gf1}), yielding the reflection phase $\theta_{r'_A}=\arg(r'_A)$. One finds 
\begin{align}
    \theta_{r_A}+\theta_{r'_A}\simeq-\pi+\sqrt{\mu}E L.
\end{align}
Consequently, the net reflection phase is obtained as
\begin{align}
    \theta&=\theta_{r}+\theta_{r'}+\theta_{r_A}+\theta_{r'_A}\nonumber\\
    &\simeq-\pi+\sqrt{\mu}EL+\frac{\lambda k_y}{2\mu^{3/2}}.
\end{align}

\section{Calculation of the retarded Green's function}

The retarded Green's function of the central irradiated region $\mathcal{G}^r(\omega)$ is calculated by the recursive Green's Function algorithm \cite{PhysRevLett.47.882,PhysRevB.75.045346}. The full Green's function is constructed by starting from the left and right initial boundaries and adding layers one by one, which uses the Dyson equation to update the Green's function as each new layer is attached. The recursive procedure is performed from left to right to obtain $\mathcal{G}_L^r(\omega)$, and from right to left to obtain $\mathcal{G}_R^r(\omega)$, thereby yielding the full response. The left and right connected Green's functions are obtained by the recursive formula. The diagonal components are given by
\begin{align}
    \mathcal{G}_{L,j}^r(\omega)&=[\omega-\check{H}_0-\check{T}^\dagger \mathcal{G}^{r}_{L,j-1}(\omega)\check{T}]^{-1},\label{eq:gl1}\\
    \mathcal{G}_{R,j}^r(\omega)&=[\omega-\check{H}_0-\check{T} \mathcal{G}^{r}_{R,j+1}(\omega)\check{T}^\dagger]^{-1},\label{eq:gr1}
\end{align}
which start from $j=1$ and $N$ in Eqs.\ (\ref{eq:gl1}) and (\ref{eq:gr1}), respectively. While the off-diagonal components are given by
\begin{gather}
    \mathcal{G}^{r}_{L,1j}(\omega)=\mathcal{G}^{r}_{L,1j-1}(\omega)\check{T}\mathcal{G}_{L,j}^r(\omega),\\
    \mathcal{G}^{r}_{R,Nj}(\omega)=\mathcal{G}^{r}_{R,Nj+1}(\omega)\check{T}^\dagger\mathcal{G}_{R,j}^r(\omega).
\end{gather}
The full Green's function can be computed by 
\begin{gather}
\mathcal{G}_{1j}^r=\mathcal{G}^r_{L,1j-1}\check{T}\mathcal{G}_{jj}^r,\\
\mathcal{G}^r_{Nj}=\mathcal{G}^{r}_{R,Nj+1}\check{T}^\dagger\mathcal{G}^r_{jj},\\
\mathcal{G}_{jj}^r(\omega)=(\omega-\check{H}_0-\check{T}^\dagger \mathcal{G}^{r}_{L,j-1}\check{T}-\check{T}\mathcal{G}^{r}_{L,j+1}\check{T}^\dagger)^{-1}.
\end{gather}

\end{document}